\documentclass[runningheads]{llncs}
\usepackage{graphicx}
\usepackage[utf8]{inputenc}
\usepackage[T1]{fontenc}
\usepackage[english]{babel}

\usepackage{dsfont}
\usepackage{amssymb}
\usepackage{amsmath,etoolbox}

\begin{document}


%
\title{Longitudinal Image Registration with Temporal-order and Subject-specificity Discrimination}
\authorrunning{Q. Yang et al.}
\titlerunning{Longitudinal image registration}

\author{Qianye Yang\inst{1} \and Yunguan Fu\inst{1,2} \and Francesco Giganti\inst{3,4} \and Nooshin Ghavami\inst{1,5} \and Qingchao Chen\inst{6} \and J. Alison Noble\inst{6} \and Tom Vercauteren\inst{5} \and Dean Barratt\inst{1} \and Yipeng Hu\inst{1,6}}


\institute{Centre for Medical Image Computing and Wellcome/EPSRC Centre for Interventional \& Surgical Sciences, University College London, London, UK 
\and 
InstaDeep, London, UK 
\and 
Division of Surgery \& Interventional Science, University College London, London, UK 
\and 
Department of Radiology, University College London Hospital NHS Foundation Trust, London, UK
\and
School of Biomedical Engineering \& Imaging Sciences, Kings College London, London, UK 
\and
Institute of Biomedical Engineering, University of Oxford, Oxford, UK \\
\email{qianye.yang.19@ucl.ac.uk}}

%
\maketitle
\begin{abstract}
Morphological analysis of longitudinal MR images plays a key role in monitoring disease progression for prostate cancer patients, who are placed under an active surveillance program. In this paper, we describe a learning-based image registration algorithm to quantify changes on regions of interest between a pair of images from the same patient, acquired at two different time points. Combining intensity-based similarity and gland segmentation as weak supervision, the population-data-trained registration networks significantly lowered the target registration errors (TREs) on holdout patient data, compared with those before registration and those from an iterative registration algorithm. Furthermore, this work provides a quantitative analysis on several longitudinal-data-sampling strategies and, in turn, we propose a novel regularisation method based on maximum mean discrepancy, between differently-sampled training image pairs. Based on 216 3D MR images from 86 patients, we report a mean TRE of 5.6 mm and show statistically significant differences between the different training data sampling strategies. 

\keywords{Medical image registration  \and Longitudinal data \and Maximum mean discrepancy.}
\end{abstract}
\section{Introduction}

Multiparametric MR (mpMR) imaging has gained increasing acceptance as a noninvasive diagnostic tool for detecting and staging prostate cancer \cite{moore2017reporting}. Active surveillance recruits patients with low-grade cancers that exhibit low-to-medium risk to long-term outcome \cite{moore2017reporting}, where mpMR imaging has been adopted to follow regions within the prostate glands and to recognise or even predict the disease progression \cite{kim2008mri}. As outlined by the panel of experts who drafted the PRECISE criteria for serial MR reporting in patients on active surveillance for prostate cancer \cite{moore2017reporting}, assessing radiological changes of morphological MR features is a key component when reporting longitudinal MR images. For individual regions of pathological interest, these morphological features include volume, shape, boundary, extension to neighbouring anatomical structures and the degree of conspicurity in these features. Pairwise medical image registration quantifies the morphological correspondence between two images, potentially providing an automated computational tool to measure these changes, without time-consuming and observer-biased manual reporting. This voxel-level analysis is particularly useful in developing imaging biomarkers, when the ground-truth of the disease progression are still under debate, as in this application, and cannot be reliably used to train an end-to-end progression classifier.

Registration algorithms designed for longitudinal images have been proposed for several other applications \cite{simpson2011longitudinal,hu2017learning,liao2012novel}, such as those utilising temporal regularisation \cite{schwartz2014locally} when applied to a data set acquired at three or more time points. Most algorithms are still based on or derived from the basic pairwise methodologies. In this work, registration of a pair of longitudinal prostate MR images from the same patient is investigated. Classical algorithms iteratively optimise a spatial transformation between two given images without using population data. For example, a fixed set of parameters in an iterative registration algorithm may work well for one patient, but unlikely to be optimal for other patients. Substantial inter-patient variation leads to the lack of common intensity patterns or structures between different prostates. Ad hoc benign foci, varying zonal anatomy and highly patient-specific pathology are frequently observed in MR images, especially within the prostate gland. This is particularly problematic for classical iterative registration algorithms, when the regions of interest, smaller and non-metastasis tumours, are confined within the varying prostate glands, such as those from the active surveillance patient cohort considered in this study. We provide such an example in the presented results using an iterative registration algorithm.

In this paper, we propose an alternative method that uses recently-introduced deep-learning-based non-iterative registration for this application. Based on the results on holdout patients, we argue that learning population statistics in patient-specific intensity patterns \cite{balakrishnan2019voxelmorph} and weak segmentation labels \cite{hu2018weakly} can overcome the difficulties due to the large inter-patient variation, for aligning intra-patient prostate MR images. In order to efficiently and effectively utilise the often limited longitudinal data for training the registration network, we compare several methods to sample the time-ordered image pairs from the same patients and those from different patients, and propose a new regularisation method to discriminate time-forward image pairs versus time-backward image pairs and/or subject-specific image pairs versus inter-subject image pairs.

We summarise the contributions in this work: 1) We developed an end-to-end deep-learning registration method tested on longitudinal images. To our knowledge, this is the first study for longitudinal MR Image registration for prostate cancer patients; 2) We present a quantitative analysis on longitudinal data sampling strategies for registration network training, with and without the new regularisation method; and 3) We report a set of rigorous results based on real clinical longitudinal data, demonstrating statistically significant differences between these sampling methods. This provides practically useful evidences for further development of the registration tools in longitudinal image analysis for active surveillance patients.

\section{Methods}

\subsection{Learning-based image registration} 

\begin{figure}[t]
\centering
\includegraphics[width=0.9\textwidth]{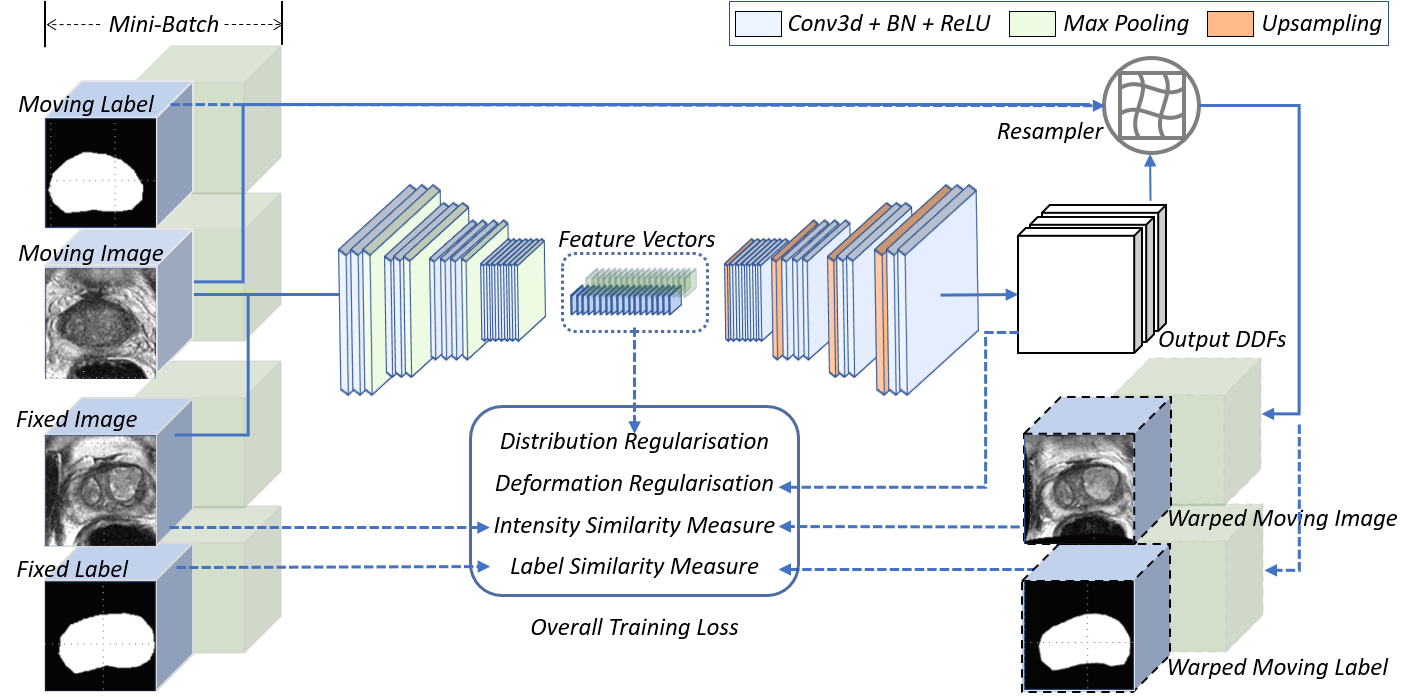}
\caption{Demonstration of the proposed image registration framework, with the dotted lines indicating the data flow only at the training.} \label{fig1}
\end{figure}

In this work, the pairwise registration paradigm based on deep learning is adopted for registering longitudinal MR image pairs. Denote $\{(x^{A}_{n}, x^{B}_{n})\}, n=1...N$, as a set of paired images to register, $x^A_n$ and $x^B_n$ being the moving- and fixed images, respectively. For each pair $n$ in the set, let a pair of corresponding prostate gland anatomical segmentation, represented with binary masks, be $(l^A_n, l^B_n)$. During training a registration network $f(\theta)$ with network parameters $\theta$, the $n$th pair of images is fed into the network to predict a dense displacement field (DDF) $\mu_n^{(\theta)}$. An overview of the network training is illustrated in Fig.~\ref{fig1}.

The training loss is comprised of three measures of the goodness-of-predicted-DDF: 1) an intensity-based similarity measure between the DDF-warped moving image $x^A_n \circ \mu_n^{(\theta)}$ and the fixed image $x^B_n$, the sum-of-square differences (SSD) in intensity values being used in this study; 2) an overlap measure between the DDF-warped moving gland segmentation $l^A_n \circ \mu_n^{(\theta)}$ and the fixed gland segmentation $l^B_n$ based on a multi-scale Dice \cite{hu2018weakly}; and 3) a deformation regularisation to penalise non-smooth DDFs using bending energy \cite{rueckert1999nonrigid}. With a minibatch gradient descent optimisation, the network weights $\theta$ are optimised by minimising the overall loss function $J(\theta)$:

\begin{equation}
    J(\theta) = \frac{1}{N}\sum\limits_{n=1}^{N}(-\alpha\cdot Dice(l^{B}_{n}, l^{A}_{n}\circ\mu_n^{(\theta)}) + \beta\cdot SSD(x^{B}_{n}, x^{A}_{n}\circ\mu_n^{(\theta)}) + \gamma\cdot\Omega_{bending}(\mu_n^{(\theta)}))
\label{eq:loss}
\end{equation}
where, $\alpha$, $\beta$ and $\gamma$ are three hyper-parameters controlling the weights between the weak supervision, the intensity similarity and the deformation regulariser. These unsupervised and weakly-supervised losses were selected based on limited experiments on a validation data set, among other options, such as cross-correlation, Jaccard and DDF gradient norms. The three weights in Eq.~\ref{eq:loss} are co-dependent with the learning rate and potentially can be reduced to two. Therefore, the fine-tuning of these hyper-parameters warrants further investigation in future studies.

\subsection{Training data distribution}

\begin{figure}
\centering
\includegraphics[width=0.8\textwidth]{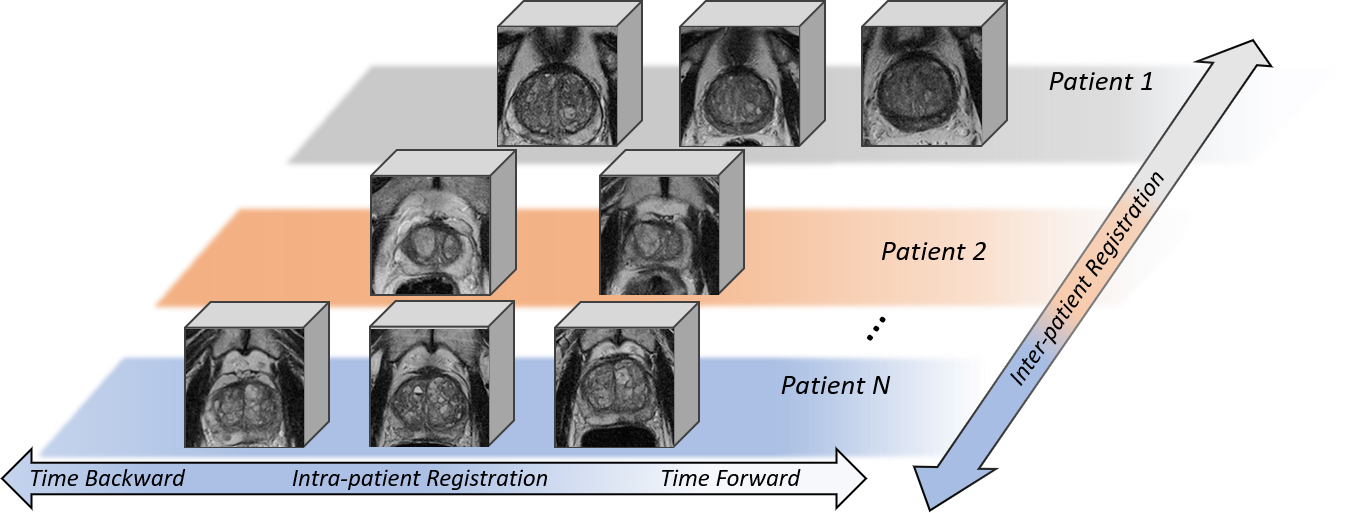}
\caption{Demonstration of a longitudinal image data set from active surveillance.} \label{fig2}
\end{figure}

\textbf{Temporal-order and subject-specificity}
Fig.~\ref{fig2} illustrates example longitudinal images from individual subjects (prostate cancer patients) at multiple visits in order of time. Without loss of generality, we aim to model the morphological changes to quantify a chronological deformation field, i.e. from a baseline T2-weighted image to a follow-up, acquired at time points $t_1$ and $t_2$, $t_1<t_2$, respectively. To train a registration network for this task, one can sample \textit{task-specific training data}, i.e. intra-subject, time-forward image pairs. Given sufficient training data, i.e. the empirical training data distribution adequately representing the population data distribution, there is little reason to add other types of image pairs, i.e. time-backward or inter-subject pairs. 

However, as is common in the field of medical image computing, the acquisition of training data is limited by various practical or clinical reasons. This leads to an empirical risk minimisation (ERM) on a sparse empirical training distribution \cite{vapnik1999overview}. ERM is known to be prone to overfitting \cite{vapnik1999overview}. Data augmentation strategies such as using affine/nonrigid transformation and the "mixup" \cite{zhang2017mixup}, in geometric transformation and intensity spaces, respectively, have been applied to overcome overfitting. The former is also applied in this work.

Particularly interesting to longitudinal pairwise image registration, image pairs with reversed temporal-order, the time-backward pairs, and those with different subject-specificity, the inter-subject pairs, can be considered as augmenting data for this task to potentially improve generalisability. This is of practical importance in training longitudinal image registration networks and is the first hypothesis to test in this study.

\textbf{Discriminating prior for regularising network training}
Furthermore, a data augmentation strategy becomes effective if the augmented empirical data distribution better represents population distribution or "aligns" empirical feature vectors with a potentially new population distribution in feature vector space \cite{tzeng2014deep}. Directly mixing the intra-/inter-subject and/or time-forward/-backward image pairs does not guarantee such alignment. To utilise the prior knowledge of the known temporal order and subject specificity, we test a new regularisation approach based on empirical maximum mean discrepancy (MMD) \cite{gretton2012kernel}. By penalising the divergence between the feature vectors generated from different empirical data distributions. As illustrated in Fig.~\ref{fig1}, an MMD square is computed over the two sets of high-dimensional feature vectors from the network encoder, $\{v_i\}_{i=1}^I$ and $\{v_j\}_{j=1}^J$, generated from different types of image pairs:

\begin{equation}
    \Omega_{mmd} = \frac{1}{I^2}\sum\limits_{i\neq i'}^{I}k_{II}(v_i,v_{i'}) - 
            \frac{2}{IJ}\sum\limits_{i,j=1}^{I,J}k_{IJ}(v_i,v_j) +
            \frac{1}{J^2}\sum\limits_{j\neq j'}^{J}k_{JJ}(v_j,v_{j'}),
\label{eq:mmd}
\end{equation}
where $k(\mu, \nu)=\exp({{-{||\mu-\nu||}^2}/{2\sigma}})$ is a Gaussian kernel with a parameter $\sigma$~\cite{bousmalis2016domain}. 
$I$ and $J$ are the sample numbers of the feature vectors, within a minibatch of size $N=I+J$. The weighted MMD regularising term in Eq.~\ref{eq:mmd} is added to the original loss in Eq.~\ref{eq:loss}: 

\begin{equation}
    J^*(\theta) = J(\theta) + \lambda\cdot\Omega_{mmd}(\{v_i\},\{v_j\}).
\label{eq:newloss}
\end{equation}
With the new loss in Eq.~\ref{eq:newloss}, we test the second hypothesis that encoding the discrimination of temporal order and subject specificity can further improve the network generalisability.

\subsection{Validation} 
\textbf{Gerneralisability on holdout data}
The patients and their data is randomly assigned into training, validation and holdout sets. The networks are developed with training and validation sets, including hyper-parameter tuning. The generalisability is measured on the holdout set using three metrics: 1) the binary Dice similarity coefficient (DSC) between the fixed label $l^{B}$ and the warped moving label $l^{A}\circ\mu$; 2) the MSE between the fixed image $x^{B}$ and the wrapped moving image $x^{A}\circ\mu$; and 3) the centroid distance between the aligned prostate glands $l^{B}$ and  $l^{A}\circ\mu$. Results from paired t-tests with a significance level of $\alpha = 0.05$ are reported when comparing these metrics. 

\textbf{Registration performance}
Also on the holdout set, pairs of corresponding anatomical and pathological landmarks are manually identified on moving and fixed images, including patient-specific fluid-filled cysts, calcification and centroids of zonal boundaries. The target registration errors (TREs) between the corresponding pairs of landmarks, from the warped moving and those from the fixed images, are computed to quantify the registration performance.  Other experiment details are provided in Section 3.

\section{Experiments}

\subsection{Data and preprocessing} 216 longitudinal prostate T2-weighted MR images were acquired from 86 patients at University College London Hospitals NHS Foundation. For each patient 2-4 images were available, with intervals between consecutive visits ranging from 11 to 28 months. All the image volumes were resampled to $0.7\times0.7\times0.7$ $mm^3$ isotropic voxels with a normalised intensity range of $[0,1]$. For computational consideration, all images were also cropped from the image center to $128\times128\times102$ voxels, such that the prostate glands are preserved. The prostate glands were manually segmented for the weak supervision in training and for validation. The images were split into 70, 6 and 10 patients for training, validation and holdout sets. 

\subsection{Network training} A previously proposed DDF-predicting encoder-decoder architecture was used \cite{hu2018weakly}, which is an adapted 3D U-Net \cite{ronneberger2015u}, with more densely connected skip layers and residual shortcuts. The network training was implemented with TensorFlow 2 \cite{abadi2016tensorflow} and made open source \url{https://github.com/DeepRegNet/DeepReg}. The Adam optimizer with an initial learning rate of $10^{-5}$ was used with the hyper-parameters $\alpha, \beta, \gamma$ and $\lambda$ empirically set to 1, 1, 50 and 0.01, respectively, via qualitative assessment on the validation set. The networks were trained on Nvidia Tesla V100 GPUs with a minibatch of 4 sets of image-label data, each containing a pair of T2 MR images and a corresponding binary mask pair of prostate gland segmentation. Each network run for 272,000 iterations, approximately 64 hours.

\subsection{Training image pair sampling} 

To test the first hypothesis in Section 2.2, three different training data sets were sampled, resulting in three networks: a network trained with only intra-subject, time-forward image pairs (IF); a second network (IF+IB) trained using all possible intra-subject pairs regardless of temporal order; and the third network (IT+IF+IB) with added inter-subject samples. All the networks were trained with the registration loss function in Eq.~\ref{eq:loss}. Generalisability and TREs were computed on all the intra-subject, time-forward image pairs from the same holdout patient data. 

To test the second hypothesis, two more networks were trained with the loss in Eq.~\ref{eq:newloss}, with respective training data sets IF+IB and IT+IF+IB. For intra-patient IF+IB pairs, two images were randomly sampled without replacement from a single patient. MMD may be sensitive to minibatch size and sample size imbalance \cite{gretton2012kernel}, a two-stage sampling was adapted to ensure every minibatch samples 2 IF and 2 IB pairs during training the IF+IB network; and samples 2 IF/IB pairs and 2 IT pairs during training the IT+IF+IB network. For comparison purposes, the same sampling was adopted when the MMD was not used. When testing the IT+IF+IB network, with or without the MMD regulariser, results were computed on all the intra-patient pairs in the holdout set.

\subsection{Comparison of networks with an iterative algorithm}

To test an iterative intensity-based registration, the widely-adopted nonrigid method using B-splines was tested on the same holdout images. For comparison purposes, the sum-of-square difference in intensity values was used as similarity measure with all other parameters set to default in the NiftyReg \cite{modat2010fast} package. The reported registration performance was to demonstrate its feasibility. Although the default configuration is unlikely to perform optimally, tuning this method is considered out of scope for the current study.

\subsection{Results}

\textbf{Sampling Strategies}
Networks with different training data sampling methods, described in Section 3.3, are summarised in Table \ref{tab1}. Adding time-backward and inter-subject image pairs in the training significantly improved the performance, both in network generalisability and registration performance. For example, the TREs decreased from 6.456$\pm$6.822 mm to 5.801$\pm$7.104 mm (p-value=0.0004), when adding time-backward data, to 5.482$\pm$5.589 mm (p-value=0.0332), when further inter-subject data was added in training. The same conclusion was also observed in DSCs and gland CDs.

\begin{table}
\centering
\caption{Registration performance with NiftyReg and networks with different sampling strategies. *See text for details including the explnation of the inferior NiftyReg result.}\label{tab1}
\resizebox{0.8\textwidth}{!}{
\begin{tabular}{c|c|c|c|c}
\hline
Methods          & DSC & CD & MSE & TRE \\
\hline
NiftyReg$^\star$        & 0.270$\pm$0.304 & 22.869$\pm$11.761mm & 0.041$\pm$0.019 & 21.147$\pm$15.841mm \\
w/o registration & 0.701$\pm$0.097 & 8.842$\pm$4.067mm & 0.051$\pm$0.090 & 10.736$\pm$7.718mm \\
IF               & 0.861$\pm$0.042 & 2.910$\pm$1.756mm & 0.049$\pm$0.097 & 6.456$\pm$6.822mm \\
IF+IB            & 0.870$\pm$0.033 & 2.257$\pm$1.503mm & 0.043$\pm$0.096 & 5.801$\pm$7.104mm \\
IT+IF+IB         & 0.885$\pm$0.024 & 2.132$\pm$0.951mm & 0.053$\pm$0.014 & 5.482$\pm$5.589mm \\
\hline
\end{tabular}}
\end{table}

\textbf{Regularisation effect}
Table~\ref{tab2} summarises the comparison between networks trained with MMD regularisation (Eq.~\ref{eq:newloss}) and without (Eq.~\ref{eq:loss}). Although improved results were consistently observed in expected DSCs and TREs, statistical significance was not found on holdout set. However, we report a statistically significant improvement on the validation set, due to the introduction of the MMD regulariser, e.g. p-values=0.016 in lowered MSEs. This was probably limited by the small holdout set and the under-optimised hyperparameters specifically for the MMD regulariser in current study.  

\begin{table}
\centering
\caption{The comparison between networks trained with and without MMD regularisation. See text for more details.}\label{tab2}
\resizebox{0.8\textwidth}{!}{
\begin{tabular}{c|c|c|c|c|c|c}
\hline
Method    & Test               & MMD      & DSC & CD & MSE & TRE \\
\hline
IF+IB     & IF                & $\times$ & 0.870$\pm$0.033 & 2.257$\pm$1.503mm & 0.043$\pm$0.096 & 5.801$\pm$7.104mm \\
IF+IB     & IF                & $\surd$  & 0.876$\pm$0.027 & 2.300$\pm$1.007mm & 0.042$\pm$0.094 & 5.847$\pm$6.360mm \\
IT+IF+IB  & IF+IB             & $\times$ & 0.890$\pm$0.019 & 1.928$\pm$0.797mm & 0.048$\pm$0.010 & 5.638$\pm$6.021mm \\
IT+IF+IB  & IF+IB             & $\surd$  & 0.893$\pm$0.023 & 1.527$\pm$0.832mm & 0.049$\pm$0.010 & 6.048$\pm$6.721mm \\
\hline
\end{tabular}}
\end{table}

\textbf{Registration performance}
As shown in Table~\ref{tab1}, the proposed registration network, with any training data sampling strategies, produced significantly lower TREs on holdout data than the TREs before registration or that from default NiftyReg algorithms. With the overall inferior NiftyReg results summarised in the table, we report that the best-performed registration from NiftyReg achieved a comparable DSC of 0.81 and a TRE of 4.75 mm, better than the network-predicted. This provides an example of highly variable registration performance from an iterative algorithm, frequently encountered in our experiment. However, a comprehensive comparison is still needed to draw further conclusions. On average, the inference time was 0.76 seconds for the registration network, compared with 50.4 seconds for NiftyReg.

\begin{figure}[t]
\includegraphics[width=\textwidth]{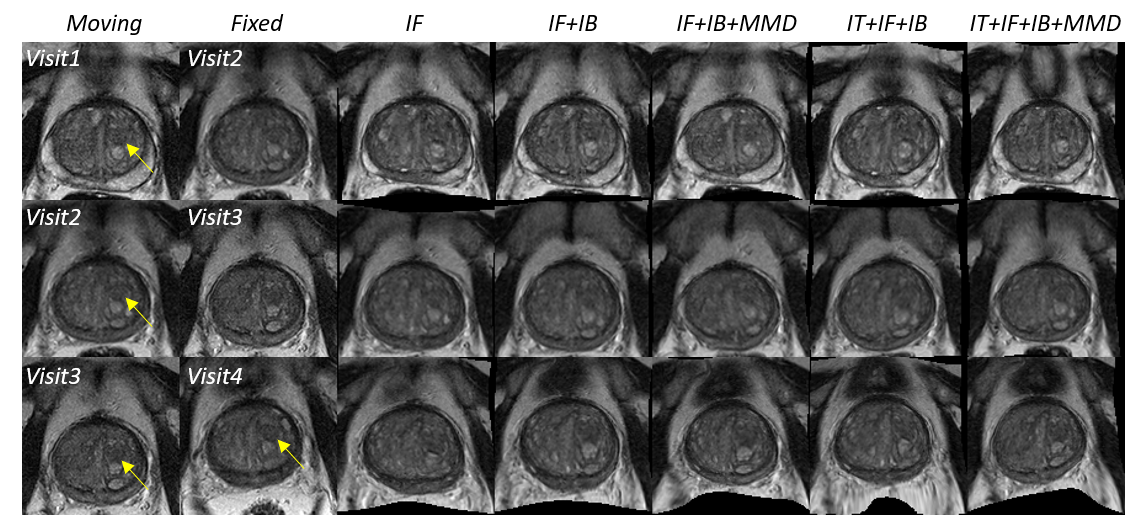}
\caption{Example registration results of a patient with 4 visits. The 1st and 2nd columns are the moving and fixed images. The remainder represent the network-warped images.}\label{fig3}
\end{figure}

\textbf{A case study for longitudinal analysis of prostate cancer}
Fig. \ref{fig3} qualitatively illustrates a 60-year-old man from active surveillance with a baseline and three follow-up visits, subject to a biopsy of Gleason 3+3. The yellow arrows indicate the evolution of a marked adenomatous area within the left transitional zone. The registration network was able to track the changes of the suspicious regions between consecutive visits, in an automated, unbiased and consistent manner over 3 years. Ongoing research is focused on analysis using the registration-quantified changes.

\section{Conclusion}
For the first time, we have developed a deep-learning-based image registration method and validated the network using clinical longitudinal data from prostate cancer active surveillance patients. We have also shown that adopting different training strategies significantly changes the network generalisability on holdout data.

\section*{Acknowledgment}
This work is supported by the Wellcome/EPSRC Centre for Interventional and Surgical Sciences (203145Z/16/Z), Centre for Medical Engineering (203148/Z/16/Z; NS/A000049/1), the EPSRC-funded UCL Centre for Doctoral Training in Intelligent, Integrated Imaging in Healthcare (i4health) (EP/S021930/1) and the Department of Health’s NIHR-fundedBiomedical Research Centre at UCLH. Francesco Giganti is funded by the UCL Graduate Research Scholarship and the Brahm Ph.D. scholarship in memory of Chris Adams.

%
%
%
\bibliographystyle{splncs04}

\begin{thebibliography}{10}
\providecommand{\url}[1]{\texttt{#1}}
\providecommand{\urlprefix}{URL }
\providecommand{\doi}[1]{https://doi.org/#1}

\bibitem{abadi2016tensorflow}
Abadi, M., Agarwal, A., Barham, P., Brevdo, E., Chen, Z., Citro, C., Corrado,
  G.S., Davis, A., Dean, J., Devin, M., et~al.: Tensorflow: Large-scale machine
  learning on heterogeneous distributed systems. arXiv preprint
  arXiv:1603.04467  (2016)

\bibitem{balakrishnan2019voxelmorph}
Balakrishnan, G., Zhao, A., Sabuncu, M.R., Guttag, J., Dalca, A.V.: Voxelmorph:
  a learning framework for deformable medical image registration. IEEE
  transactions on medical imaging  \textbf{38}(8),  1788--1800 (2019)

\bibitem{bousmalis2016domain}
Bousmalis, K., Trigeorgis, G., Silberman, N., Krishnan, D., Erhan, D.: Domain
  separation networks. In: Advances in neural information processing systems.
  pp. 343--351 (2016)

\bibitem{gretton2012kernel}
Gretton, A., Borgwardt, K.M., Rasch, M.J., Sch{\"o}lkopf, B., Smola, A.: A
  kernel two-sample test. Journal of Machine Learning Research
  \textbf{13}(Mar),  723--773 (2012)

\bibitem{hu2017learning}
Hu, S., Wei, L., Gao, Y., Guo, Y., Wu, G., Shen, D.: Learning-based deformable
  image registration for infant mr images in the first year of life. Medical
  physics  \textbf{44}(1),  158--170 (2017)

\bibitem{hu2018weakly}
Hu, Y., Modat, M., Gibson, E., Li, W., Ghavami, N., Bonmati, E., Wang, G.,
  Bandula, S., Moore, C.M., Emberton, M., et~al.: Weakly-supervised
  convolutional neural networks for multimodal image registration. Medical
  image analysis  \textbf{49},  1--13 (2018)

\bibitem{kim2008mri}
Kim, C.K., Park, B.K., Lee, H.M., Kim, S.S., Kim, E.: Mri techniques for
  prediction of local tumor progression after high-intensity focused ultrasonic
  ablation of prostate cancer. American Journal of Roentgenology
  \textbf{190}(5),  1180--1186 (2008)

\bibitem{liao2012novel}
Liao, S., Jia, H., Wu, G., Shen, D., Initiative, A.D.N., et~al.: A novel
  framework for longitudinal atlas construction with groupwise registration of
  subject image sequences. NeuroImage  \textbf{59}(2),  1275--1289 (2012)

\bibitem{modat2010fast}
Modat, M., Ridgway, G.R., Taylor, Z.A., Lehmann, M., Barnes, J., Hawkes, D.J.,
  Fox, N.C., Ourselin, S.: Fast free-form deformation using graphics processing
  units. Computer methods and programs in biomedicine  \textbf{98}(3),
  278--284 (2010)

\bibitem{moore2017reporting}
Moore, C.M., Giganti, F., Albertsen, P., Allen, C., Bangma, C., Briganti, A.,
  Carroll, P., Haider, M., Kasivisvanathan, V., Kirkham, A., et~al.: Reporting
  magnetic resonance imaging in men on active surveillance for prostate cancer:
  the precise recommendations—a report of a european school of oncology task
  force. European urology  \textbf{71}(4),  648--655 (2017)

\bibitem{ronneberger2015u}
Ronneberger, O., Fischer, P., Brox, T.: U-net: Convolutional networks for
  biomedical image segmentation. In: International Conference on Medical image
  computing and computer-assisted intervention. pp. 234--241. Springer (2015)

\bibitem{rueckert1999nonrigid}
Rueckert, D., Sonoda, L.I., Hayes, C., Hill, D.L., Leach, M.O., Hawkes, D.J.:
  Nonrigid registration using free-form deformations: application to breast mr
  images. IEEE transactions on medical imaging  \textbf{18}(8),  712--721
  (1999)

\bibitem{schwartz2014locally}
Schwartz, E., Jakab, A., Kasprian, G., Z{\"o}llei, L., Langs, G.: A locally
  linear method for enforcing temporal smoothness in serial image registration.
  In: International Workshop on Spatio-temporal Image Analysis for Longitudinal
  and Time-Series Image Data. pp. 13--24. Springer (2014)

\bibitem{simpson2011longitudinal}
Simpson, I.J., Woolrich, M., Groves, A.R., Schnabel, J.A.: Longitudinal brain
  mri analysis with uncertain registration. In: International Conference on
  Medical Image Computing and Computer-Assisted Intervention. pp. 647--654.
  Springer (2011)

\bibitem{tzeng2014deep}
Tzeng, E., Hoffman, J., Zhang, N., Saenko, K., Darrell, T.: Deep domain
  confusion: Maximizing for domain invariance. arXiv preprint arXiv:1412.3474
  (2014)

\bibitem{vapnik1999overview}
Vapnik, V.N.: An overview of statistical learning theory. IEEE transactions on
  neural networks  \textbf{10}(5),  988--999 (1999)

\bibitem{zhang2017mixup}
Zhang, H., Cisse, M., Dauphin, Y.N., Lopez-Paz, D.: mixup: Beyond empirical
  risk minimization. arXiv preprint arXiv:1710.09412  (2017)

\end{thebibliography}

\end{document}